\documentclass{tcibook}
\usepackage{fancyhea}
\usepackage{work}
\usepackage{bm}       
\usepackage{graphicx}
\usepackage{hyperref}      
\usepackage{multirow}
\usepackage{lineno}

\newcommand{\nc}{\newcommand}  

\def\beq{\begin{equation}}
\def\eeq#1{\label{#1}\end{equation}}
\def\eeqn{\end{equation}}

\newenvironment{Eqnarray}%
   {\arraycolsep 0.14em\begin{eqnarray}}{\end{eqnarray}}
\def\beqa{\begin{Eqnarray}}
\def\eeqa#1{\label{#1}\end{Eqnarray}}
\def\eeqan{\end{Eqnarray}}

\nc{\ra}{\rightarrow}  
\nc{\slsh}{\slash\hspace*{-0.22cm}}
\def\Re{{\cal R \mskip-4mu \lower.1ex \hbox{\it e}\,}}
\def\Im{{\cal I \mskip-5mu \lower.1ex \hbox{\it m}\,}}

\nc{\vev}[1]{ \left\langle {#1} \right\rangle }
\nc{\bra}[1]{ \langle {#1} | }
\nc{\ket}[1]{ | {#1} \rangle }
\nc{\fb}{\,{\rm fb}^{-1}}
\nc{\ev}{{\rm eV}}
\nc{\kev}{{\rm keV}}
\nc{\Mev}{{\rm MeV}}
\nc{\gev}{{\rm GeV}}
\nc{\tev}{{\rm TeV}}
\nc{\mev}{{\rm MeV}}

\def\del{\partial}
\def\Dslash{\not{\hbox{\kern-4pt $D$}}}
\def\dslash{\not{\hbox{\kern-2pt $\del$}}}
\def\pslash{\not{\hbox{\kern-2pt $p$}}}
\def\ETmiss{ \not{\hbox{\kern-4pt $E$}}_T }

\def\msb{{\bar{\ssstyle M \kern -1pt S}}}

\begin{document}

\def\bibname{References}
\bibliographystyle{plain}

\raggedbottom

\pagenumbering{roman}

\parindent=0pt
\parskip=8pt
\setlength{\evensidemargin}{0pt}
\setlength{\oddsidemargin}{0pt}
\setlength{\marginparsep}{0.0in}
\setlength{\marginparwidth}{0.0in}
\marginparpush=0pt


\pagenumbering{arabic}

\renewcommand{\chapname}{chap:intro_}
\renewcommand{\chapterdir}{.}
\renewcommand{\arraystretch}{1.25}
\addtolength{\arraycolsep}{-3pt}

\thispagestyle{empty}
\begin{centering}
\vfill

{\Huge\bf Planning the Future of U.S. Particle Physics}

{\Large \bf Report of the 2013 Community Summer Study}

\vfill

{\Huge \bf Chapter 5: The Future of \\ U.S. Particle Theory}

\vspace*{2.0cm}
{\Large \bf Convener: M. Dine}

\vfill

{\large  Study Conveners: M. Bardeen, W. Barletta, L.~A.~T.~Bauerdick, R. Brock,
D.~Cronin-Hennessy, M.~Demarteau, M.~Dine, J.~L. Feng, M. Gilchriese,
S. Gottlieb, J.~L.~Hewett, R. Lipton, H.~Nicholson, M.~E. Peskin,
S. Ritz, I.~Shipsey, H. Weerts}\\
\vspace{1cm}

{\large Division of Particles and Fields Officers in 2013:
J.~L. Rosner (chair), 
I. Shipsey (chair-elect), 
N. Hadley (vice-chair),
P. Ramond (past chair)}\\
\vspace{1cm}

{\large Editorial Committee:
R.~H. Bernstein,
N. Graf,
P. McBride,
M.~E. Peskin,
J.~L. Rosner,
N.~Varelas,
K. Yurkewicz}

\vfill

\end{centering}

\pagenumbering{roman}

\newpage
\mbox{\null}

\vspace{3.0cm}

{\Large \bf Authors of Chapter 5:}

\vspace{2.0cm}

{\bf M. Dine}, 
K. Babu,
C. Csaki,
S. Dawson,
L. Dixon,
S. Gottlieb,
J. Harvey,
D. Whiteson

 \tableofcontents

\newpage

\mbox{\null}

\newpage

\pagenumbering{arabic}



\setcounter{chapter}{4}

\chapter{The Future of U.S. Particle Theory}
\label{chap:Theory}

\begin{center}\begin{boldmath}



\begin{center}

\begin{large} {\bf Convener: M. Dine} \end{large}

K. Babu,
C. Csaki,
S. Dawson,
L. Dixon,
S. Gottlieb,
J. Harvey,
D. Whiteson

\end{center}


\end{boldmath}\end{center}

\section{Introduction}

Theoretical physics has played a crucial role in particle physics
since its earliest days.  Interpretation and synthesis of a broad
range of experimental results (phenomenology), progress in quantum
field theory (``formal'' theory)\footnote{While the term ``formal" is often used, we prefer to refer
to such work as ``research on foundational questions" or ``field theory, quantum gravity, and string theory", and will do so
in the rest of this document.}, recognition of the role of
symmetries (model building), and invention of new calculational methods
(perturbative techniques and lattice QCD) were all crucial in
developing the Standard Model (SM).  Theorists have been the drivers in
asking the questions which lead beyond the SM, including:
the origin of the hierarchy between the scales of the weak and
gravitational interactions, the physics of flavor, the origin of
neutrino masses, the particle nature of dark matter, inflationary
cosmology, baryogenesis, resolving the tension between quantum mechanics
and gravity, and identification and explanation of dark energy.

Since the Second World War, the U.S. has been a world
leader in theoretical particle physics.  This has remained the case in
recent years, despite the move of the Energy Frontier to CERN.  This
leadership results from a combination of university-based and national
lab-based research, supported principally by the Department of Energy
and the National Science Foundation.

The DPF Theory Panel was formed with the goal of
understanding both the scientific problems and opportunities of the next
decade, as well as the challenges involved in sustaining a first-class
program in the U.S.
Specifically, the panel's charge included:
\begin{enumerate}
\item    Enumerate areas of opportunity in particle physics theory research in order to set forth a vision for theoretical high energy physics for the next several years.
\item    Establish a range of funding needs for individual PI's to sustain an effective program
(students, postdocs, travel, summer salary, equipment needs).
\item    Examine roles and relative funding levels of university and national lab theory groups. This includes examination of Comparative Review processes for these two categories of research groups.
\item    Consider if suitable mechanisms are in place to assure funding of young researchers.
 \end{enumerate}
To address these questions, we solicited comments and suggestions
from the community, held town meetings at the BNL and KITP pre-Minneapolis workshops
leading up to the main meeting in Minneapolis, and held two parallel
sessions and a plenary session in Minneapolis.

Our basic conclusions and recommendations can be simply summarized.  The U.S.
should maintain a vigorous research effort in theoretical particle physics,
ranging from perturbative and non-perturbative QCD studies, to collider
phenomenology, to model building, cosmology, and research in
foundational areas.
We will provide more detailed recommendations in section \ref{recommendations}.

Theorists and theory straddle the Energy, Intensity and Cosmic Frontiers throughout the international
particle physics program, and have an
important role to play in the developing international particle physics program.
Theorists continue to provide the overarching intellectual framework
for these programs.  They also more directly help to enunciate
the physics cases for future experimental facilities, and to define
many of the analyses to be performed at the LHC and beyond.

The SM is likely
complete, in the sense that it is a consistent theory up to very high energies.  It is now the background to all of our experimental
explorations. There are many questions beyond the SM, and the field now confronts new and serious challenges.  In going beyond the SM, we are entering an environment with both high potential
rewards and high risks.  Theorists will continue to propose possible new
phenomena relevant to all three Frontiers --- Energy, Intensity and
Cosmic --- which will help guide future
experimental studies and will respond to them.  They will extend the structures of
quantum field theory and quantum gravity, providing clues as to the
possible underlying laws of nature, and they will continue to propose
explanations for the hierarchy of energy scales, the origin of dark matter
and similar mysteries, with
implications for existing and proposed experiments. As they have
for decades, they will  provide critical input to analysis of experimental data.

Trends in funding endanger this vital enterprise.  The U.S. funding agencies anticipate
significant future reductions in funding for theoretical physics, and these
are likely to harm the depth, breadth, and world
leadership of this program.  In FY 2013, support for particle physics in the NSF was cut by 10-12\%.  The DOE is facing a declining budget and is
increasing the fraction of its budget devoted to new experimental and accelerator
projects at the expense of research funding. Unlike many elements of the experimental effort, theoretical
research does not lend itself to ``project" designation, so the impact of this shift
on theory is more pronounced.  The consequences of these cuts
may be amplified, if applied uniformly to research groups.  This is particularly true for postdocs, as hiring a research associate requires some minimum funding level, and many groups are likely to find their funds fall below this minimum.
Additional cuts to graduate
student support will lead to shrinking numbers of individuals admitted to
study particle theory, as well as a longer time to Ph.D. for those
who remain.  This will have a significant negative impact on our
field for years to come.   Preserving postdoctoral support at the
present level is essential for the health of the field, since postdocs are the drivers of many challenging problems, and also the future
leaders of the field. The recent adoption of a comparative review
process in the Department of Energy has the potential to allow more targeted
cuts, allowing for some control of the numbers of postdocs and students, but even then, serious harm will occur if current budget trends continue.

\section{Particle physics and theoretical particle physics today}

The past two years have seen a major triumph of theoretical and
experimental physics working hand in hand.  A
scalar particle has been discovered in the mass range expected for the Higgs boson from analyses
of precision electroweak data.  This is an extraordinary experimental
accomplishment.  But this success also rests on
our exquisite understanding of QCD, and our ability to predict,
for the simplest Higgs theory, the production rate and the decay
branching ratios with great precision.  Early indications are that
this particle is in fact the Higgs predicted by the simplest version of
the SM, where simple has a precise meaning: it is the minimal
number of degrees of freedom consistent with the symmetries of the strong,
weak, and electromagnetic interactions, and the principles of quantum
mechanics, locality, and special relativity.

Theorists are playing an important role in upcoming tests of the SM
interpretation of the Higgs boson, which will be central to the
continuing LHC program and to a possible International Linear Collider
(ILC).  Not only are they providing the required SM
calculations for production and decay rates, new ways to test
the properties of the Higgs boson, and alternative models against
which these measurements can be tested, but they have a framework
in which to quantify any would-be discrepancies between the simplest
Higgs theory and experiment (using the methods of ``effective field theory").

On a slightly longer time scale, we have seen, over the the past two decades, many other successes of the
SM.  Among the most striking in recent years
has been the experimental verification that the CKM phase explains
the observed CP violation in the $K$ and $B$ meson systems.
This is a triumph for the SM and for experimental
ingenuity.  However, it also reflects the development, over the past three
decades, of a spectrum of theoretical tools, including: the general framework for
weak interaction phenomenology employing the operator product expansion; the
recognition of the incisiveness of time-dependent CP asymmetries;
novel methods for understanding heavy quark systems; and an
extraordinary increase in the ability to compute real
physical quantities from lattice gauge theory.  Indeed, the progress
in lattice gauge theory over the past decade has been astounding,
including, for example, calculations of the hadron spectrum yielding
precision measurements of light quark masses and computation of decay
amplitudes necessary for the precise extraction of CKM parameters
from data.

Another success for theory (together with experiment)
has been the verification of SM predictions with amazing precision in collider
experiments.  Experiments at LEP and SLC
confirmed the electroweak couplings with great accuracy and tested
QCD in a clean environment.  Theorists and experimentalists
have combined data from HERA and the Tevatron to provide precise knowledge of the momentum distributions
of quarks and gluons in the proton.  This knowledge, together with improvements
in the theory of hadron collider processes, has enabled SM tests
and measurements
at the Tevatron and the LHC with a precision
that was unthinkable even a decade ago.  The $W$ mass measurement at the Tevatron, and the tests of the SM electroweak vector boson production rates and kinematic distributions, as well as the top quark pair-production cross sections at the Tevatron and the LHC, are just a few examples of this progress.

The past decade has also seen confirmation of the existence of
neutrino masses and the measurement of neutrino oscillation
parameters.  Theorists have played an important role in every aspect of these
developments.  They first laid the theoretical foundation for
understanding neutrino oscillations, in vacuum and including
matter effects, and provided natural frameworks for explaining
the extremely small values of the neutrino masses.  Theorists also provided a detailed understanding of
issues arising in the detection of these phenomena using neutrinos
from the Sun, the atmosphere, nuclear reactors, and accelerators.

Our seemingly complete understanding of physics, down to distance
scales of order $10^{-17}$ centimeters, has brought other questions
about the laws of nature into sharp focus.  Again, theory plays a
crucial role in delineating the questions and in suggesting possible
answers.  Among the questions which theorists have helped to identify
and sharpen are:
\begin{enumerate}\item
What is the origin of the great disparity in the energy scales associated
with the weak and gravitational forces?  This is the hierarchy
problem.  It has two pieces:  1)  why is there such a large
disparity 2)   the problem of fine tuning: any new energy threshold much above the masses of the $W$ and $Z$ bosons, such as
the Planck scale or unification scale, tends to destabilize the Higgs boson mass through quantum corrections.
\item Where do the parameters of the SM originate?
\item Do the strong and electroweak forces unify at some energy scale?
\item
Why is the strong interaction CP conserving?  Is this accounted for by an axion field, and does this axion
constitute some or all of the dark matter?
\item The quarks and leptons present many mysteries.  Why are there repetitive generations?
What accounts for the hierarchical structure of the masses and mixings of
the quarks and charged leptons?
\item  The discovery of neutrino mass has raised new questions.
What is the energy scale associated with the generation of
neutrino mass?  Are neutrinos their own anti-particles?
\item The observed CP violation in the SM is insufficient to
account for the baryon asymmetry of the Universe.  What phenomena might
account for this?  Might they be accessible to experiments at the Energy
or Intensity Frontiers?
\item What is the identity of the dark matter which makes up 25\% of the
energy density of the Universe?
\item What is the origin of the dark energy which makes up 70\% of the
energy density? Why is it just becoming
important at the present epoch of the Universe?
\item	What caused the inflationary epoch, and how did the Universe
end up in its current state?
\item What is the nature of the quantum theory of gravitation?
\item From what set of principles or structures do the laws of nature
originate?

\end{enumerate}

Theorists are vigorously considering all of these questions.  Some of
them point to particular energy scales and types of experiments.
Others questions in the list are more speculative.  These questions straddle --- as do the interests of most
theorists --- the Energy, Intensity, and Cosmic Frontiers.  Proposals for physics beyond the SM include:
\begin{enumerate}
\item[{\bf---}] Supersymmetry, a possible new symmetry of nature relating fermions
and bosons, to understand the hierarchy between the Planck
scale and the weak scale.  In many realizations that theorists have considered,
one might have expected its discovery in the first run at the LHC.  Still, it remains one of the more plausible explanations,
and is the subject of continued experimental and theoretical study.
\item[{\bf---}] Composite Higgs models, technicolor, and Randall-Sundrum models.
These provide alternative possible explanations of the hierarchy problem,
and are the subject of ongoing experimental searches.
\item[\bf---] Dark matter candidates.  Weakly interacting massive particles (WIMPs)
are natural in supersymmetry and several other theoretical structures;
axions were invented to understand the strong CP problem.  These are both
topics of ongoing theoretical work and extensive experimental searches.
\item[{\bf---}] String theory and other ideas for a quantum theory of gravity.  String theory in particular provides a promising model for the unification of
gravity and the other forces in a consistent quantum mechanical framework.
It has also provided new tools for addressing problems in quantum field theory
and in disparate areas of physics including heavy ion physics and
condensed matter physics.  It has suggested new principles (holography) and
inspired ideas for particle phenomenology and physics beyond the
SM. It has also inspired the invention of powerful techniques
for computing scattering amplitudes.
\item[{\bf---}] Leptogenesis:  This is an attractive paradigm for explaining the baryon asymmetry of
the Universe, which has an intimate connection with
the origin of neutrino masses.  Plausible indirect evidence for this
mechanism would be the discovery of CP violation in the neutrino sector,
the subject of tests in forthcoming long-baseline experiments.  Other ideas for baryogenesis
have different potential consequences.
\end{enumerate}
In addition to raising questions, theorists have developed powerful perturbative and
non-perturbative techniques for
performing calculations essential for understanding collider experiments.
Theoretical precision that was previously viewed as impossible is now routine.
This precision has been, and will remain, crucial in both understanding SM
physics and uncovering evidence of physics beyond the SM.
This includes understanding
Tevatron and LHC data, as well as results from BES III and Belle-II.

Upcoming results from the LHC or from experiments at any of the frontiers in the next
decade might provide answers to some of these questions.
For example, the hierarchy (or ``naturalness'') problem provides the principal argument that new physics should
appear at the TeV-scale, and it has inspired a range of proposals for
physics beyond the SM.

The results of the first 25~fb$^{-1}$ of integrated luminosity
collected at the LHC have placed many of these questions in a new
perspective.  The current data are consistent with the
SM, with a single Higgs doublet and a corresponding particle
with mass $126$ GeV.  Many ideas about the hierarchy problem predict the existence of
new colored particles to cancel large contributions to the Higgs mass from the
top quark; there are now strong limits on such particles
with masses below a TeV.
Many specific proposals for new physics at the TeV-scale have been
severely constrained by LHC searches, and the paradigm of naturalness
has come under increasing scrutiny. However, there remains significant space to explore.

Answering theoretical questions will require experiments across the Frontiers.  Intensity Frontier experiments and precision
measurements at the Energy Frontier may provide
evidence for new physics at slightly higher energy scales.  Cosmic Frontier experiments may
yield a WIMP candidate or further constrain this paradigm.
For other questions, experimental input is
likely to be limited for a long while, and theorists will try to put
together answers, at first tentative, that combine experimental knowledge
with theoretical insight.  Aspects of the frontier classification scheme, and its relevance to theory, will be discussed in Section \ref{frontiers}
below.

\section{Prospects for advances in theory}

During the various Snowmass workshops, and especially in Minneapolis,
we asked theorists with a broad range of research interests to outline recent developments and to
enumerate areas of opportunity likely to witness significant progress
over the next decade. We present some highlights of these
discussions.  As with most theoretical ideas, several topics discussed
below have overlapping boundaries.   This list does not encompass all active areas of theoretical particle
physics, but does represent a significant fraction of current activity.  It is important to note that theorists
not only cross the frontiers, but also the areas of activity we enumerate below.

Here we consider five broad areas: phenomenology, field theory calculational
methods, model building,  astrophysics, and cosmology, and string theory,
quantum gravity and foundational questions.  These are broad topics; within
phenomenology, for example, we consider collider phenomenology, electroweak
physics, neutrino physics, heavy quark physics and additional topics; in field
theory, we include lattice gauge theory as well as perturbative and
semiclassical methods; model building includes models of flavor,
supersymmetry, grand unified theories (GUTs), and large or warped dimensions.

\subsection{Phenomenology}

Particle phenomenology plays the crucial role of linking experiments
with the various aspects of theory, including model building, perturbative and lattice QCD, and
more foundational issues.  Over the past decades it has been central to
the extraction of the
parameters of the fundamental Lagrangian from experimental data. It played a vital
role in the success of the Tevatron and LHC programs,
leading to precision determination of the top quark properties and the
discovery of the Higgs boson. The successful operation of the $B$ factories
led to the confirmation of the Cabibbo-Kobayashi-Maskawa paradigm
for quark mixing and CP violation, and the ever more precise
determination of the parameters of the flavor sector.
The phenomenology community has proposed many new kinematic variables now used on a
daily basis by researchers in the LHC collaborations.  Collider phenomenologists have
also catalyzed an important new area of study at hadron colliders:
jet substructure and various other jet and event properties, such as
$N$-jettiness and $N$-subjettiness, have led to new experimental
strategies for searching for pure-jet decays of heavy particles
and of the Higgs boson. Phenomenology has also played a crucial role in the
direct and indirect dark matter detection experiments.   In particular,
theorists
working on dark matter have
proposed that the proper framework of analyzing direct detection
experiments is in terms of the couplings of a low-energy effective
theory, broadening the possible interpretations of these
experiments.  Phenomenology will remain critical to the interpretation of continuing and upcoming experiments and
for developing plans for the future.

\subsubsection{Flavor physics} 

The study of flavor physics was an integral part of the development of the SM. The existence of the charm quark was predicted based
on the suppressed decay $K_L \rightarrow \mu^+ \mu^-$.  The fact that
$\nu_\mu$ is a distinct state from $\nu_e$ was inferred
from the absence of $\mu \rightarrow e\gamma$ decay,
and the top quark was predicted to be heavy from the
measurement of $B_d-\overline{B}_d$ mixing.  In the quark sector,
the Cabibbo-Kobayashi-Maskawa mixing and CP violation paradigm was
confirmed with data from BaBar, Belle, and the Tevatron.  These measurements
are quite impressive, but experimental and theoretical
uncertainties still leave some room for contributions to processes in the $B$ system.  Theory has
played a crucial role here, beginning with the idea of three
generations of quarks, to heavy quark effective field theory and
lattice QCD, which provided important form factors with a few percent
uncertainty.

Proposals for physics beyond the SM are strongly constrained by flavor physics,
whether or not they address directly the big questions of the subject.  The
scale of new physics that contains generic flavor violations is constrained by
present data on rare processes in the muon, kaon, $B$ and $D$-meson systems to
be greater than about $10^4$ to $10^5$ TeV. For example, the spectrum of
supersymmetric particles must exhibit special features.  For example, the
squark masses might be nearly the same (degeneracy), or they might be
approximately diagonal in a basis in which the fermion masses are (alignment),
if they are within reach of currently conceivable experiments.
In the near future, LHCb and Belle-II will test
non-standard theories of quark flavor with higher precision, and theorists will be engaged
with analysis and interpretation of the results.    Electric
dipole moment limits on atoms, nuclei, the neutron and the electron
severely constrain new sources of TeV-scale CP violation. Theoretical
ideas which attempt to explain the pattern of fermion masses and mixings
also typically predict new sources of
flavor violation.  The scale of flavor dynamics could be low, comparable to the TeV scale,  or much higher.
If it is not too high, experiments can test these ideas. The theory of flavor is relatively clean in processes such as $\mu \rightarrow e \gamma$
decay and coherent $\mu \rightarrow e$ conversion in nuclei (once the source of flavor violation is
specified), but not so clean at the required level of accuracy in muon $g-2$ for
which there is currently a 3.6 $\sigma$ discrepancy between theory and
experiment. For the next round of the $g-2$ experiment, lattice
calculations may decrease the theoretical uncertainty so
that experimental results can be better compared to the SM
prediction.

\subsubsection{Neutrino physics}

Neutrino theory sits at the intersection
of particle physics, nuclear physics, astrophysics and cosmology, and
as such provides great opportunities and interesting challenges.
These are especially relevant today, given that a significant part of the U.S.
experimental program over the next decade is likely to be focused on
measurement of neutrino properties.

The experimental neutrino physics program has had tremendous
success over the past fifteen years, beginning with the discovery of
neutrino oscillations in solar and atmospheric neutrinos.
Theory has played an important role in each step of these
developments, from the calculation of neutrino fluxes from the Sun and
from the atmosphere, to the recognition
of the importance of matter effects including MSW resonances in neutrino
propagation, to the calculation of various
neutrino cross sections, to elucidating mechanisms for generating
small neutrino masses.

This has focussed the field
on measurement and understanding of the neutrino masses and mixings (the PMNS matrix), as
well as fundamental issues such as  whether neutrinos are there own antiparticles (Majorana particles), the scales of neutrino
mass generation, and the possible role of neutrinos
in the creation of the asymmetry between matter and antimatter (leptogenesis).
The recent
measurement of one of the neutrino mixing parameters, $\theta_{13}$, in reactor-  and accelerator-based
experiments is a major step in pinning down the PMNS matrix.     Discovery of CP violation in neutrino
oscillations, as anticipated in forthcoming long baseline neutrino
experiments in the U.S. and abroad, could provide indirect hints for
leptogenesis as a plausible mechanism for generating the baryon
asymmetry of the Universe.    To unravel answers to the
fundamental questions, theorists must undertake the time-consuming
task of combining different experimental results, each with its own
uncertainties, nontrivial correlations, and parameter degeneracies.
Often the needed calculations, such as neutrino-nucleus cross sections,
require a strong background in nuclear physics.  The three-flavor
oscillation paradigm needs to be tested by over-constraining
parameters, which in itself is a time-consuming
endeavor, requiring years of dedicated efforts.  Currently, the U.S.-based theory community working on such topics is relatively small.
There is an ongoing effort to enhance this community, led by a Neutrino
Theory Task Force.  This Task Force proposes new initiatives to attract more researchers into this
area, in collaboration with experimentalists and nuclear theorists.

\subsection{Field theory calculational methods}

\subsubsection{Perturbative and effective field theory methods}

In a variety of circumstances, it is possible to extract important results from
experiment only through very precise theoretical predictions.
For example, a more precise understanding of the physics
of heavy hadrons has been enabled by several breakthroughs:
first, the operator product expansion for weak transitions;
and later, the successive development of heavy quark effective theory,
nonrelativistic QCD, and most recently, soft collinear effective theory
(SCET).  These developments all represent the construction of effective
field theories (EFTs) that arise from a separation of physical scales.
They have had great practical payoff for the experimental study of $B$
mesons, charmonia, and bottomonia.

Monte Carlo event generators, such as Isajet, Pythia, Herwig, Sherpa,
Alpgen and Madgraph, were developed by the theory community and have
become indispensable tools for experimentalists to simulate events
at all high-energy colliders, but particularly hadron colliders.
There has also been enormous progress in computing cross sections relevant
to both backgrounds and signals at colliders, at higher order in the
strong coupling constant, $\alpha_s$, and even including electroweak corrections.
Next-to-leading order (NLO) QCD predictions for the LHC are now routine
and automated in programs such as MCFM, Rocket, CutTools, GoSam and BlackHat,
based in part on an improved theoretical understanding of loop amplitudes.
The next order (NNLO QCD) is now in sight for a variety of LHC processes
as well, where it will make even more precise experimental tests possible.
This progress at fixed order in the coupling has been accompanied by
a greatly improved understanding of how to resum large logarithmic
corrections:  either analytically using, for example, SCET; or
within Monte Carlo simulations by merging fixed-order results with
parton showers.

Precise parton distributions are essential for all
LHC physics, but it was only possible to obtain these once NLO and especially NNLO
computations were combined with high precision data from HERA and elsewhere.
QCD has truly come into its own as a precise theory for hadronic collisions.
One also cannot overstate the importance of precision electroweak theory.
The triumph of precision electroweak measurements in predicting
the Higgs mass would not have come about if the quantum corrections
to the weak mixing angle and weak boson masses had not been known
to two loop order.  With only the one-loop terms, the inferred Higgs boson
mass would have been about 500 GeV higher than where the Higgs was
found.  Even better theoretical precision will be needed to match the
experimental capabilities of an ILC running at the $Z$ pole.
Despite all this progress in precision applications of
continuum quantum field theory, there is still plenty of room for new
theoretical developments, both those with practical applications to
experiments, and those that enlarge our understanding of quantum field
theory and of the structure of physical law.

\subsubsection{Lattice QCD}

Lattice QCD is our main tool for understanding non-perturbative aspects of QCD.
Numerical evaluation of hadronic matrix elements is crucial to progress in many areas of particle
physics.  The last decade has witnessed an enormous increase in the power of lattice
methods.
Lattice QCD has enabled the computation of weak matrix elements at the
percent level, along with
precise determination of the QCD coupling $\alpha_s$ and the quark
masses.  Computations of the decay constants of the $D$ and $D_s$ mesons
provide striking examples of recent progress.  Other applications include:  1)  studies
of condensed matter physics 2)
work by the Columbia and Edinburgh lattice QCD groups
in collaboration with IBM  that resulted in the Blue Gene
series of supercomputers 3) the MILC collaboration
code has been used as a benchmark for many
computer purchases.   Breakthroughs are expected
over the coming years, including:
determination
of $m_b$ and $ \alpha_s$ to $0.25\%$ or better (important for
precise calculation of Higgs decay modes at the proposed ILC);
further calculations of weak matrix elements and form
factors needed for tests of the CKM paradigm;
calculation of the hadronic contributions to
the muon magnetic moment, important for the interpretation
of the upcoming $g-2$ measurement;
nucleon matrix elements needed for prediction of nucleon decay and $n-\bar n$ mixing;
form factors for quasi-elastic neutrino-nucleon scattering;
calculations of strongly-coupled models of physics beyond the SM;
and lattice calculations of supersymmetric models.
Calculations are increasingly done with up and down
quark masses at the average of their physical values.  Future calculations will
likely include dynamical electromagnetism and isospin breaking.
Lattice gauge theory needs increased support at universities as well as laboratories, particularly
in order to ensure training of a future generation of students.

\subsection{Model building}

Model building connects theory to phenomenology. Model builders
discover new mechanisms, new phenomena, and inspire experiments and
new experimental analyses.  Here we distinguish TeV-scale model building, and model building involving higher energy scales, such as GUTs, models
for neutrino mass and flavor, and string model building.  Much of TeV model building has been driven by
efforts to solve the hierarchy problem, but an important motivation is enumerating possible signals
for experiments in the complicated environments of the Tevatron and LHC.  This will continue to be an important driver
of theoretical work in this area in the coming decades.  Higher scale model building yields possible solutions to
many of our outstanding questions.  In some cases, it has led to model building of relevance to lower energy
physics, and/or has driven experiments which have shed light in other areas.

\subsubsection{TeV-scale model building}

For TeV model building,
the idea of supersymmetry, which can
potentially resolve the hierarchy problem, has been an important driver of experiments.
Supersymmetry predicts an array of new particles, often with properties which are
a priori known, and this has led to a variety of
experimental searches looking directly for these new particles.  Many of the searches for dark matter via direct
detection have focused on the lightest supersymmetric particle, which is a
natural candidate for dark matter.  The theory of large extra spatial
dimensions stimulated short-distance tests of Newton's laws.  The axion
hypothesis led to
microwave cavity experiments and is pointing towards new types of
experiments.  Warped extra-dimensional models led to searches for the
Kaluza-Klein gluon, which in turn was one of the inspirations for
new techniques related to jet substructure.
New types of models for dark matter led to new classes of
experiments searching for dark sectors.
Models of leptogenesis provide a very strong
motivation to map out the parameters of the neutrino sector in great
detail, especially the CP properties.

In the LHC era, in which
trigger bandwidth is critical and backgrounds are large, it is important
to have models that cover as many signatures as possible.
For example, $R$-parity-violating supersymmetric models inspire new
searches for colored states decaying without missing energy.
Split supersymmetry suggests quasi-stable gluinos and leads to searches
for out-of-time decays.  Other models lead to quasi-stable charged
particles or displaced vertices.
Having a
large number of exotic models significantly expands the range of search
strategies at the
LHC experiments, reducing the likelihood of missing
critical signals of new phenomena. This synergy between experiments, phenomenology and
model-building can work in both directions.  Hints of new physics
suggested by experiments, for example the muon $g-2$ anomaly, have
greatly influenced model building as well as phenomenology.

\subsubsection{Flavor model building}

The origin of the quark, lepton, and in recent years
the neutrino mass matrices are among the great mysteries
in particle theories.  Theorists have explored a variety of ideas for understanding the hierarchies
observed in the quark and charged lepton mass matrices.  These have included, but are hardly
limited to, flavor symmetries with small breakings, warped and extra dimension models,
and composite and technicolor models.  The evolving determination of neutrino masses and mixings poses
challenges for many of these ideas, but at the same time, neutrino model building ties closely to possible mechanisms
for the origin of the matter-antimatter asymmetry.

\subsubsection{Grand unified theories}

GUTs are proposals to unify the strong, weak, and
electromagnetic interactions into a single force.
They also unify quarks, leptons, anti-quarks and anti-leptons of each
family into common multiplets. Such an arrangement would explain the
coexistence of quarks with leptons, their quantum numbers, and the
quantization of electric charge. With the assumption of low energy
supersymmetry, the three gauge couplings of the SM are
found to unify rather precisely when extrapolated to high energies at
a scale $M_X \approx 2 \times 10^{16}$ GeV, which may be argued to be a piece of
indirect evidence for grand unification and for supersymmetry.

This unification is the principal direct evidence for GUTs.  But the study of these theories has had
other impacts.
GUTs stimulated proton decay experiments, which helped solve the solar neutrino problem, and
led to the discovery of atmospheric neutrino oscillations, and to the detection of neutrinos from a supernova.

Even without supersymmetry,
the gauge couplings can unify if the GUT permits an
intermediate symmetry, as happens in SO(10).  SO(10) theories also
predicted the existence of the right-handed neutrino, which plays an
essential role in the generation of small neutrino masses via the
seesaw mechanism.  Unification of matter into common multiplets
implies that baryon number is not conserved, and that the proton should
decay.  In the context of low energy supersymmetry, these theories
predict that the decays $p \rightarrow \overline{\nu} K^+$ and $p
\rightarrow e^+\pi^0$ should occur with rates that are likely to be
within reach of the next-generation proton decay search experiments.

Proton decay searches probe
distance scales of order $10^{-30}$ cm, something not possible by any
other means.  If discovered, proton decay would be a landmark
discovery in science.

\subsubsection{String theory model building}

There are many challenges in connecting string theory to the real world, but consideration of string models has profoundly influenced ideas for
particle physics models.  Some of these ideas will be mentioned below in Section \ref{stringtheory}, but further examples
include ideas for understanding light Higgs bosons, theorems demonstrating the absence of continuous global
symmetries, the role of discrete symmetries, ideas for the origin of repetitive generations, a natural setting for the Peccei-Quinn solution
of the strong CP problem, candidate fields and mechanisms for inflationary cosmology, and alternatives to the conventional hot Big Bang
theory at early times.  We anticipate further progress in this field, exploiting theoretical developments and responding to experimental discoveries and exclusions.

\subsection{Astroparticle physics and cosmology}

The combination of a cosmological constant with a distribution
of cold dark matter particles ($\Lambda$CDM) after an initial
period of inflationary expansion is now considered the
SM of cosmology.  This model has provided a simple
explanation of diverse physical phenomena in the Universe.  However,
even if this simple picture continues to be borne out,
there is much still to explain, including the reason the cosmological
constant has the value it does, the origin of cosmological density
perturbations, and the nature of dark matter.
Theorists have proposed a range of dark matter candidates; indeed,
the need for cold dark matter
has become a criterion for model building and particle phenomenology.
Astrophysical and cosmological theory also plays a key role in the
interpretation of a wide variety of current and future measurements,
including dark matter experiments, massive galaxy surveys,
gravitational lensing observations, and cosmic microwave background data.
The hypothesis of inflation came from particle theory, emerging from
ideas about grand unification and monopoles.  It is now a central
part of the Cosmic Frontier and of the Theory program.  Recent data from WMAP and Planck
have further constrained possible inflationary models.  Given
the limited amount of data we are likely to have, theoretical
ideas will be critical if there is ever to be a single, compelling
model underpinning inflation.  Such ideas might come, for example, from string theory, or from
connecting the degrees of freedom of inflation with those
accessible to experiments (e.g., excitations in large extra dimensions
or associated with supersymmetry breaking).
%
%
%
\subsection{String theory/quantum gravity/foundational questions}
\label{stringtheory}

The SM developed, in
significant part, from theorists delving into fundamental questions in
field theory. These included the demands of renormalizability and unitarity and
the related constraint of freedom from anomalies.  Its development and full
understanding required that theorists master previously unfamiliar topics
in mathematics such as group theory,
topology and the theory of fiber bundles.
String theory grew out of efforts to understand the phenomenology of the strong
interactions, and later blossomed into a unified approach to quantum
mechanics and gravity. In particular, the resurgence
of interest in  string theory in 1984 followed the
discovery of anomaly-free string theories that could incorporate the chiral gauge structure
of the SM.  Since then string theory has been a major source of
new ideas in particle physics.  It has provided insights into questions of unification, the
strong CP problem, black hole physics, supersymmetry, the possibility of
flat and warped extra dimensions, and much more.  It has had an important indirect impact on particle physics by inspiring new computational approaches to ordinary perturbation theory.

One of the most important recent developments in string theory is the ``AdS/CFT" correspondence, or gauge/string duality.  This is the startling
observation that a quantum gravity theory in Anti-deSitter space is equivalent to a conformal field theory at the
boundary of the space.  This idea has provided a fundamental
new tool for the study of strongly interacting field theories. As such it has provided a new method of
studying non-perturbative QCD, has motivated new computations in lattice gauge theory, has found
important applications to heavy ion physics, where it was used to predict the viscosity to entropy ratio
of the quark-gluon plasma, and is now being widely applied to problems in condensed matter physics.

As a theory of quantum gravity, string theory also has close ties to general relativity and there has been
a fruitful interplay between string-theoretic methods and more conventional methods in the study of
the properties of black holes.  Supersymmetry, alone and in conjunction with techniques developed in string theory,
continues to be a powerful tool for unraveling the dynamics of
strongly-interacting gauge theories. String theory practitioners have also
contributed to advances in our understanding of cosmology (e.g., the
systematic study of non-gaussianity), models of flavor physics, and
much more. String theory and supersymmetry have also had a broad impact
in pure mathematics in areas ranging from algebraic geometry to number theory.
It is likely that the ideas and techniques to which string theory has led will
be critical to resolving many of the questions we have about nature at the deepest level.

\section{Particle theory, the Frontier paradigm and cross-disciplinary research}
\label{frontiers}

It is worth stressing again that theoretical particle physics, and most theorists, transcend the
frontiers of physics.  QCD, the theory of the
strong interactions, is an essential element in both the Energy and Intensity Frontiers.  QCD computations
are important for the Cosmic Frontier, for example in evaluating
cross sections for the direct detection of dark matter.
Perturbative and lattice QCD are essential for understanding experiments in both
areas.  Furthermore, from a theoretical point of view, the physics of heavy
quarks straddles the Energy and Intensity Frontiers, and has received
contributions from theorists with a broad range of interests.  In
building models to describe short-distance phenomena, theorists will
consider the collider signatures of their theories, the constraints
from and predictions for flavor changing processes, and possible
electric dipole moments.  They will almost invariably ask whether
their models possess a dark matter candidate, and if so, what sorts of
experiments might detect it.  They will consider cosmological
issues, including creation of the dark matter and the baryon
asymmetry, mechanisms for inflation and possible distortions of the microwave background, and the
like.   String theorists have pointed out the possibility
of alternative cosmologies, in which the Universe was not hot in the
past, and provided a variety of frameworks for inflation.  They have
also suggested that axion decay constants might be larger than
conventionally supposed, inspiring new cosmologies and exploration
of new search strategies.  String theorists and others working on more
foundational questions have provided tools for scientists working on heavy
ion physics (and other areas outside of traditional particle
physics), as well as increasing our knowledge of field theory in ways
which have proven useful for perturbative and non-perturbative QCD, and for
those building models which go beyond the SM.

In addition to overlapping several frontiers, research in particle theory influences, and is influenced by,
other areas of physics.  The DOE and NSF have historically supported a broad science program, and, as
indicated above, this has had significant scientific rewards.
The question has arisen whether the traditional breadth of scope of the
U.S. program in elementary particle theory should be narrowed in fiscally
challenging times.  There has been
much discussion, in particular, about subjects like AdS/CMT, the  application of gauge/string duality to condensed matter systems,
 with some questioning why the DOE, in particular,
should support this activity.  We would respond that there is a long history of fruitful cross-pollination of ideas between
condensed matter physics and particle theory. For example, the BCS theory of superconductivity, which we now
recognize as the first example of spontaneous breaking of a gauge symmetry, provided direct
inspiration for Nambu's Nobel Prize winning work on chiral symmetry breaking in strong interactions. The renormalization group and its
revolutionary implications for our understanding of quantum field theory
had its origins in particle theory.  It was then combined with ideas of
universality and scaling and applied to understand phase transitions in
condensed matter physics; insights gained from these applications
then filtered back into particle physics. In the modern era, insights gained from the study of topological objects
like solitons and instantons have had an impact on both condensed matter and particle physics. The recent progress in the condensed matter physics of topological insulators
owes much to the understanding of anomalies gained by particle theorists. Recently these connections have expanded through
gauge/string duality to include both general relativity and string theory.

There has also been increasing interaction between particle theory and
areas of pure mathematics, an area of research sometimes
referred to as ``physical mathematics."   For example, there
are burgeoning connections between number theory, geometry and the
mathematical structure of scattering amplitudes.
There has also been a resurgence of interest in the formal structure of
supersymmetric gauge theories and their application to areas of mathematics,
including knot theory and the structure of low-dimensional manifolds.
Dualities in string theory have found a direct connection to elements of the Langlands correspondence, one of the main drivers of research in mathematics.
These areas are further from DOE's traditional purview (NSF is able, in some cases, to share such efforts between its Math and Physics Divisions), but may also lead
to new tools and ideas with applications to particle physics.  They are being
pursued by some of the most talented members of the younger generation.

We see many reasons that such activities
should be supported. First, the questions which drive our field may require entirely new
concepts and methods.  As Dirac famously said in
the paper in which he introduced the magnetic monopole,  ``The steady progress of physics requires for its theoretical formulation a
mathematics that gets continually more advanced."  Second, these subjects have
many ideas and techniques in common that  often find application in areas far from their original source.
Finally, we must note that these connections are at the cutting edge of research in the more foundational aspects of the subject and attract
many outstanding young theorists.  We will lose these people to other areas or other countries if their research is not supported.

Theoretical physics is in many ways a unified field in which the U.S. has been the
dominant force in the world.  Our universities and advanced
graduate programs still lead the world but are under severe threat from funding cuts.  We currently attract many of the top students
from abroad.  These students not only receive a world-class education, but also become familiar with our culture of scientific excellence.
Other countries understand the power of educating the best minds in the world.
The panel believes it would be a great loss if the U.S. gave up this role.

\section{The roles of national laboratories and universities}

A significant share of research in theoretical physics is performed in National Laboratories, with a roughly
equal level of funding for labs and universities.  The number of PI's in the labs is significantly smaller than in universities, and the
funding balance results, in part, from the fact that lab PI's work throughout the year on research (and professional
service), without obligations to teach and to university service.

The U.S national labs are critical players in the national theory effort.
They support excellent theoretical programs,
which provide leadership in research on QCD and Higgs physics
relevant to collider physics, and essential support for ongoing experimental
analyses and planning for future facilities.  Essentially all
of the U.S. expertise in parton shower Monte Carlos resides at the labs,
and they have a strong focus on lattice and perturbative QCD as well.

Concerns about declining funding
have led some to suggest an examination of  the balance of lab and
university funding, and we have discussed this among ourselves and
with the agencies.
These individuals have pointed out that, while the national labs
represent about 20\% of PI's nationwide and a comparable fraction of total output (as measured by publications, impact factors and similar
measures) they represent about 48\% of the total expenditure of DOE theory
funds.  This results from the fact that universities, as knowledge producers
as well as educational institutions, pay for much of the research time of their
faculty.  This balance has served physics education and the field of particle physics well for many decades.

Still, declining funding has serious implications for research in labs and universities, both for the present cadre of researchers
and research activity, and for future hiring in the field.
We encourage HEPAP to establish a subcommittee to look at the complex question
of university/lab balance.

\section{Sustainability: theory funding}
\label{theoryfunding}

The U.S. has sustained a vigorous program in theoretical
particle physics for many decades.
Support for this effort has come principally from
the DOE and the NSF, with
modest additional funding from private sources.
We have had extensive conversations with representatives of the
DOE and NSF, and have been gratified by their strong appreciation of the value of theory.
This is, however, a challenging period for funding of all
aspects of particle physics; we have focused on a number of issues particular to theory.
We are especially concerned, as we will describe in greater detail
in this Section, with the consequences of anticipated future cutbacks for postdoctoral
fellows and graduate students in national laboratories and universities, and for hiring of new faculty in
theoretical particle physics at U.S. universities.

DOE and NSF funding has
supported the following activities central to the theory effort:
\begin{enumerate}
\item {\bf Training of students.}
U.S. universities still lead the world in attracting the strongest students in many
subfields of theoretical particle physics.
Besides direct support for graduate student research through grants
to PI's, support has also been provided through the DOE graduate
student fellowship and the Fermilab visiting student program, and the LHC-TI graduate student fellowship funded by the NSF.
The Theoretical Advanced Study Institute (TASI) summer school (for many years held at UC Boulder) has provided,
since 1984, a thorough training for advanced graduate students
in modern particle theory concepts and methods.  It is an extremely valuable component of the U.S. particle theory funding portfolio, and is the most highly regarded and competitive summer school anywhere in the world.
Many TASI students have gone on to become leading researchers
in particle theory.  The federally supported training of graduate students
directly supports the future of particle theory in the U.S.
However, the effects of training students are broader than simply their impact on particle physics.
Many students eventually leave the field,
typically put their training to use in the ``knowledge economy'' of
the U.S., whether it is in other fields of research, in teaching at
undergraduate institutions, Silicon Valley, Wall Street, or other
technical endeavors.
\item {\bf Salaries of postdoctoral fellows.}
Postdocs enable much of the particle theory research performed
in the U.S. and constitute the future leadership
of the field.
While a substantial fraction go on to positions in universities and national labs,
others move out of particle physics and contribute
to the broader technical work force in the U.S.
\item {\bf Travel to conferences and workshops and for collaboration.}
Even in the internet era, face-to-face communication between theorists
is essential for propagating ideas to others and for developing collaborations.
\item {\bf Summer salary for investigators.}
Summer salary frees faculty from teaching responsibilities in summer
months and facilitates concentrated research time.    The case for hiring faculty in particle theory
at U.S. universities is driven in part by the recognition that they
play an important role in nationally supported research.  In addition to taking on teaching
in the summer to supplement their incomes,
faculty
lacking summer salary are often assigned increased teaching loads
during the academic year.
\item {\bf Theoretical physics in the national laboratories.}
Traditionally, the DOE has supported strong
research groups at its national laboratories that have performed
theoretical research closely related to experiments in particle physics.  These theory groups
have performed original research and trained postdocs and students.
There has been a strong focus
on relevance to, and support for, experimental activities, particularly
those related to on-site accelerator-based facilities, but also including
more general planning of large national and international facilities.
While there is at present only one single purpose national laboratory dedicated to particle physics (Fermilab),
the other labs maintain research groups in experimental
particle physics working on the LHC detectors, on dark matter experiments worldwide,
on the LSST, and other experiments.  These efforts benefit from the presence of laboratory theory groups.
\item{\bf Institutes running extended workshops.}  A very important component contributing to  the vitality and success of U.S. particle theory is the set of workshops run by the Aspen Center for Physics (ACP) and the Kavli Institute for Theoretical Physics
(KITP at UC Santa Barbara), both  partially funded by the NSF.  Their workshops focus on a particular topic of current interest. An Aspen workshop typically runs for 3--5 weeks, with around 30 participants per week, while a KITP workshop might run for a few weeks or a whole semester.  These programs have become major venues for intense exchanges of ideas, and canalization of new projects. KITP programs provided the impetus for much of the early progress in AdS/CFT, for the development of new methods for perturbative computations, and for numerous other activities across the field.  The ACP hosted the discovery of anomaly cancellation in string theory, a major impetus
to the field, and was the place where the original idea for the e-print archive arXiv was formulated.   Numerous projects in model building, neutrino physics and collider phenomenology had their genesis at the Aspen Center.  These institutes have become so successful that many other countries (and subfields) have put similar institutions in place.  Examples include the Galileo Galilei Institute (GGI) in Florence, Italy, the Munich Institute for Astro-and Particle Physics (MIAPP) in Munich, Germany, the Mainz Institute for Theoretical Physics (MITP) in Mainz, Germany, the Kavli Institute for Theoretical Physics China (KITPC) in Beijing, China, as well as the Institute for Nuclear Theory (INT) in Seattle funded by the DOE Nuclear Physics program, and the Simons Center for Mathematical Physics at Stony Brook. In order to maintain the ACP's and the KITP's status as the world-leading institutes for extended workshops,  it is very important to keep funding these institutes at appropriate levels, allowing future generations of theorists to take advantage of these excellent resources.
\end{enumerate}
This formula for supporting particle theory in the U.S. has been extremely successful.  The theory program in
the U.S. has arguably been second to none for many decades.
With a changing funding climate, this model is at risk.  Funding
for DOE-funded groups is expected to decline several percent per year
for the next several years; the NSF has seen a severe decline, as we have mentioned,
in 2013 alone.  This has a number of implications:
\begin{enumerate}
\item {\bf Decline in support for graduate students.}
In most university groups, particle theory students are currently serving as teaching assistants (TAs)  during significantly more
than half
 of the academic terms in which they
are enrolled.  These TA duties lengthen the time required to
complete a Ph.D. thesis, with detrimental effects on education and careers.
\item {\bf Decline in support for postdoctoral fellows.}
Postdoctoral fellows play a vital role in particle theory groups, and the postdoctoral period is a crucial one in a research career.   To prove that he or she qualifies to become a permanent member of the research community, a postdoctoral researcher must be open to new directions, invent or develop new concepts or methods that push the state of the art in some area, and evolve a unique, personal outlook within the research enterprise.
These experiences often guide the researcher throughout his or her whole career.
Developing a distinct professional identity is a far greater challenge than that of completing a PhD project with
one's adviser.  Many of those who succeed at the first level
cannot achieve this one; others emerge because the brilliance of their independent work. With the current and anticipated future funding cuts there is a possibility that the number of postdoctoral positions might shrink drastically in a short period of time. Any appreciable decrease in the pool of postdocs would have a significant negative impact on the future of the field.
\item {\bf Severe restrictions on travel.}
Travel budgets at universities and the national labs are shrinking substantially. Senior researchers are often forced to choose between travel essential for their research and travel by postdocs and students. It should be noted that, in addition, the approval process at national labs has become increasingly cumbersome and often is now hindering healthy exchanges of scientific ideas.
\item{\bf Elimination of Laboratory Visitor Programs, Fermilab and DOE student fellowships.}
One of the roles of theory groups at national labs is to be the national centers of scientific interaction. However, the recent elimination of visitor programs at national labs makes this impossible. This is in stark contrast with CERN, where the theory group is truly the center of European particle theory, mainly due to its visitor program and focused theory institutes. The very successful Fermilab visiting student program (``Fermilab Fellowship for Theoretical Physics")
faces an uncertain future beyond the 2013-2014 academic year.  Similarly, the DOE graduate student fellowship program has been suspended.
\item{\bf Cap on university summer salaries.}
The DOE has instituted a \$15K/month cap per investigator for summer salaries, which was also implemented this year in the NSF theory program. This effects only more highly paid senior faculty, and is preferable to further cuts in postdoc and student support.  Significant cuts beyond this level could have negative effects on more junior faculty and overall productivity.
\item{\bf The move towards a higher fraction of projects within the DOE has led to  disproportionately large cuts in the theory budget.}
At the recommendation of its 2010 Committee of Visitors, DOE HEP is moving towards increasing the fraction of projects in its funding portfolio to a level close to 20\%. This is clearly the correct general decision: a vibrant HEP experimental program has to have a large fraction of the budget committed to projects, rather than to general research. However, the implementation of this move has also included the theory budget, which has no projects. This has led to a significantly higher effective cut in support for theory than in some other areas.
\end{enumerate}
Each of these items has significant detrimental implications for the
future of the field.  Students make significant contributions
 to the research enterprise, both through their own work and as a
significant source of new ideas and stimulation for their mentors; at the same time, it is the training of students
that insures the future health and vibrancy of the field.  Decreasing the number of postdocs will have a significant negative
effect on research productivity, while again shrinking the pool of
future investigators.  Postdocs, particularly, bring fresh ideas to
the research enterprise, and typically bring skills and a willingness
to learn new techniques and methods. We are concerned that even modest
funding decreases will mean that many groups will drop below threshold
to sustain postdoctoral positions, so that a 10\% decrease in
funding over one to two years could translate into a much larger
decrease in the number of postdocs.  In the case of graduate students,
in the past, decreases in funding could be ameliorated with increased
TA responsibilities.  While this would slow student progress and
decrease productivity, it would at least provide a means of support.
However, particularly in public institutions, the number of TA
positions has been decreasing in many cases, closing this option.

This past year, the DOE instituted caps on summer salaries, and the
NSF is following suit.  We agree that this is preferable to further
cuts in student and postdoctoral support, but it should be noted that
still lower caps will have implications for research productivity,
particularly if they reach the level of  junior faculty (assistant or associate professor salaries).  Many researchers may have to supplement their
income with further teaching or other responsibilities in the summers.

\section{Recommendations}
\label{recommendations}

The DOE, in the context of the P5 process, has referred explicitly to
the role of theory.  In particular, P5 is to address ``Fundamental
questions for the field and how to inform/connect the Frontiers
framework $\ldots$ Input from the theory community will be especially important in this
area.'' We have outlined above our reasoning in support of these
statements, describing both the historical role of theory and what we
view as its crucial importance to the field going forward.  In
addition to enumerating what we view as the important contributions of
theorists for the national and international particle physics efforts,
we have also listed some specific areas where we see the field as
particularly vulnerable.
%
\begin{itemize}
\item Seemingly modest cuts in theory funding are likely to lead to much
larger declines in the numbers of postdoctoral fellows and students
funded.  This has implications both for research productivity and for
sustaining an outstanding theory program in the U.S.  This
includes impact on the historic ability of the U.S. to
attract outstanding talent from abroad.
\item Shrinking numbers and size of theory grants will have impact on
university hiring and teaching loads  in theoretical particle physics, as well as obvious
impacts for national labs.
\end{itemize}

Our recommendations to deal with these issues can be summarized as follows:
\begin{enumerate}
\item   It is important to maintain the vitality and international competitiveness of theory programs both
at universities and national laboratories.
\item  The move to extract funds from research for projects should treat theory differently from other areas, as the
damage to the program is more severe.  At the same time, a possible class of projects for theory is proposed below.
\item  A project category for theory, within DOE, could be the existence of theory {\it networks}, modeled loosely on such networks in Europe.  A proposal would come from multiple institutions in response to a DOE solicitation to establish such networks, open to all subfields.  We envision the scale of such networks as including of order five postdoctoral fellows, funds for travel between institutions and resources particular to the project. The projects would have a duration of order five years.  They would have well-defined deliverables.  Proposals would be peer-reviewed, preferably within a panel structure. Actual topics would emerge from the community, but one might imagine subjects such as neutrino physics (in this context, a proposal might have a component aimed at DOE Nuclear Physics), computations related to collider physics, or topics in more foundational areas (the ``String Vacuum Project" of a few years ago, funded by the NSF, is a possible model).
\item The breadth of the topics and research areas supported in particle theory should be maintained. The successful formula of funding the best and most interesting research should not be changed. While in the experimental program there is obviously a need for prioritizing based on an agency's missions, this is not the case for particle theory. We advocate that programmatic considerations in funding decisions should be kept at a minimal level.  It is important not to limit the scope of high-quality
theoretical research that is being performed, even if it appears
to cross traditional funding agency boundaries.
\item HEPAP should examine the question
of the balance of resources between laboratory and university theory groups.  Both have been vital, historically, to the theory effort in the U.S.  Both are vulnerable in the likely long term funding
environment.
\item
A target level of average support for funded researchers in university groups should include 1/2 postdoc per PI and 1/2 student (equivalent to fully supported) per PI.  It should include
travel funds adequate to attend two major meetings per year per PI, as well as one trip for postdocs and attendance every other year
by students at summer schools or appropriate conferences.  It should include two months of PI summer salary.  Levels of support will vary among groups based on assessments by
referees, comparative review panels, and agency officials.\item Support for graduate student research should be increased.
Ideally, particle theory students would be supported for three months
during the summer, and for 50\% of the terms during the academic year.
\item  PI summer salary caps should not be lowered significantly below their current levels except to protect postdoc and graduate student support, and only with
attention to impacts on research productivity, particularly of more junior PI's (assistant and associate professors).
%
\end{enumerate}

In broad brush, our recommendations are similar to the conclusions of the recent update of the European Strategy for Particle Physics
on the importance of
theoretical physics:
``Theory is a strong driver of particle physics and provides essential
input to experiments, witness the major role played by theory in the recent
discovery of the Higgs boson, from the foundations of the SM
to detailed calculations guiding the experimental searches. Europe should
support a diverse, vibrant theoretical physics programme, ranging from
abstract to applied topics, in close collaboration with experiments and
extending to neighbouring fields such as astroparticle physics and cosmology.
Such support should extend also to high-performance computing and software
development."

Similar support for our recommendations can be found in the statement, issued
by a large group of experimental particle physicists, principally from U.S.
institutions: ``We, the undersigned
experimental high-energy physicists, believe that a strong experimental
high-energy physics program requires a vibrant theoretical physics community
in the U.S.
Over the last 50 years and more, the theoretical and experimental
high-energy physics communities have supported and inspired each other.
Not only are predictive tools developed by the theoretical community
essential for the experimental endeavor, but the combination of
experimental discovery and theoretical interpretation has led to
revolutions in our understanding of the nature of matter. An experimental
program which probes today's outstanding questions is strongest when
stimulated by new ideas from the theory community.
When faced with difficult budgeting choices, we urge policy makers to
protect the strength of the theoretical high-energy physics community
and the balance between the experimental and theoretical programs.''
Signatures can be viewed at
\url{http://amanda.uci.edu/~daniel/theory\_letter.php}

In the following subsections,
we have included some topics which would be appropriate for consideration by the DOE Committee of Visitors, as well as some further information on the Snowmass Theory Panel.

\subsection{ Suggestions for agency committees of visitors}

There are several issues that have come to this panel's attention, that are more appropriately addressed by the Committee of Visitors at the DOE and the NSF. These include:
\begin{enumerate}
\item NSF panels are FACA panels.  The DOE Comparative Review panels are not.  The DOE COV should examine whether this limits the role of the panels, and consider whether a change in procedures is needed.
\item  The DOE Comparative Review process was established, in part, to assure that awards were based on the quality of proposals and not on historical factors, as sometimes was the case in the past.  Because of the challenging funding climate, it appears that in some cases, decisions about awards were made based on historical levels
of support.  The COV should examine whether, in fact, awards are in line with the quality of proposals, or whether there are still distortions due to past funding levels.
\item  The COV should examine policies regarding overlapping proposals.  For example, it should be possible to propose the same research for a DOE Career award and for a
regular DOE grant (clearly with the understanding that only one  would be funded).  We understand that the agencies have made efforts to allow such overlap, but in the past, in some cases, agency policies have led to weakened proposals, and a review would be appropriate.
\item It appears to have become standard practice in DOE not to fund proposals from first year assistant professors. The DOE COV could consider whether this serves the best interest of the field.
\end{enumerate}

\subsection{The Theory Panel:  additional information}

\noindent {\bf  Theory Panel Web Site}
The theory panel web page can be viewed at:\\
\url{http://www.snowmass2013.org/tiki-index.php?page=Theory+Panel}

\noindent{\bf Theory Plenary Presentations During the Community Summer Study}

\begin{enumerate}
\item	QCD (Kirill Melnikov)
\item	Lattice Gauge Theory (Steve Gottlieb)
\item	Neutrino Physics (Andr\'{e} de Gouv\'{e}a)
\item	Phenomenology (Tim Tait)
\item	Model Building (Ann Nelson)
\item	Cosmology (Jonathan Feng)
\item	Field Theory and String Theory (David Gross)
\item	The view from the NSF (Keith Dienes)
\item  The view from the DOE (Simona Rolli)
\end{enumerate}

Slides from most of the talks can be viewed at:  \\
\url{https://indico.fnal.gov/conferenceTimeTable.py?confId=6890\#20130804.detailed}



\end{document}